\documentclass[conference]{IEEEtran}
\usepackage{cite}
\usepackage{amsmath,amssymb,amsfonts}
\usepackage{algorithm} 
\usepackage{algpseudocode}
\usepackage{graphicx}
\usepackage{textcomp}
\usepackage{xcolor}
\usepackage{soul}
\usepackage{comment}

\usepackage{hyperref}
\usepackage{lmodern}
\usepackage{comment}
\usepackage{hyperref}
\DeclareGraphicsExtensions{.eps}
\begin{document}

\title{A UAV-Aided Digital Twin Framework for IoT Networks with High Accuracy and Synchronization} 

\author{
    \IEEEauthorblockN{Ghofran Khalaf}
    \IEEEauthorblockA{
    \textit{American University of Beirut} \\
     Beirut, Lebanon \\
    gyk03@mail.aub.edu}
    \and
    \IEEEauthorblockN{May Itani}
    \IEEEauthorblockA{
    \textit{Beirut Arab University}\\
    Beirut, Lebanon \\
    ma.itani@bau.edu.lb}
    \and
    \IEEEauthorblockN{Sanaa Sharafeddine}
    \IEEEauthorblockA{
    \textit{American University of Beirut} \\
     Beirut, Lebanon \\
    ss30@aub.edu.lb}
}
\maketitle

\begin{abstract} 
Digital Twin (DT) serves as a foundational technology for Smart Manufacturing and Industry 5.0 as it enhances data-driven decision-making in industrial environments. With the continued growth of its core technologies, including the Internet of Things (IoT), artificial intelligence (AI), Big Data and data analytics, and edge computing, DT has witnessed a significant increase in industrial applications, helping the industry become more sustainable, smart, and adaptable. Hence, DT technology has emerged as a promising link between the physical and virtual worlds, enabling simulation, prediction, and real-time performance optimization. This work aims to explore the development of a high-fidelity digital twin framework, focusing on synchronization and accuracy between physical and digital systems to enhance data-driven decision making. To achieve this, we deploy several stationary UAVs in optimized locations to collect data from industrial IoT devices, which were used to monitor multiple physical entities and perform computations to evaluate their status. We consider a practical setup in which multiple IoT devices may monitor a single physical entity, and as a result, the measurements are combined and processed together to determine the status of the physical entity. The resulting status updates are subsequently uploaded from the UAVs to the base station, where the DT resides. In this work, we consider a novel metric based on the Age of Information (AoI), coined as the Age of Digital Twin (AoDT), to reflect the status freshness of the digital twin. Factoring AoDT in the problem formulation ensures that the DT reliably mirrors the physical system with high accuracy and synchronization. We formulate a mixed-integer non-convex program to maximize the total amount of data collected from all IoT devices while ensuring a constrained AoDT. Using successive convex approximation, we solve the problem and conduct extensive simulations then compare the results with baseline approaches to demonstrate the effectiveness of the proposed solution.
\end{abstract}



\section{Introduction}\label{intro}
With emerging 5G networks that ensure instantaneous connectivity to billions of Internet of Things devices, the demand for accurate and fast data updates is increasing to facilitate remote monitoring and control~\cite{Pan22, Hak24}. This presents significant challenges for network operators in providing dynamic adaptation to customer needs. To overcome these challenges, many industries are using digital twins to improve operational efficiency and build accurate maintenance strategies. 
A recent study estimates that 85\% of IoT industries are expected to incorporate digital twinning in the near future~\cite{Xu19}. 
A digital twin is a virtual representation of a physical system, process, or object designed to accurately reflect the physical entity through synchronization and real-time data updates. It uses simulation, machine learning, and reasoning to make better decisions~\cite{Wu21,IETFDatatracker}. Digital Twin platforms provide networks with improved automation, resilience testing, full life cycle operation, and infrastructure maintenance \cite{IETFDatatracker}.

Since a large amount of real-time data is required to establish a high-fidelity digital twin \cite{IETFDatatracker}, IoT devices often fail to deliver their real-time measurements to the base station due to limited transmit power and processing capacity. Utilizing unmanned aerial vehicles (UAVs) to collect and process data from IoT devices offers a reliable and affordable wireless connectivity solution and is widely adopted in similar problem setup~\cite{Raj23, basnayaka2024freshness,gao2023aoi,zhang2024aoi}. The large amount of real-time data enforces adequate synchronization and accuracy between the physical and digital twins. On one hand, synchronization can be improved by reducing the time gap between the current status of the monitored physical entity and its digital counterpart, referred to as the Age of Digital Twin (AoDT). AoDT was first introduced in ~\cite{Ita24} and is based on the AoI, which reflects information freshness and was first introduced in 2010 ~\cite{sun2022age}. Thus, AoDT measures status freshness to determine whether the digital twin accurately mirrors the current condition of a given physical entity. On the other hand, a higher level of accuracy can be achieved by maximizing the amount of data collected. However, jointly enhancing synchronization and accuracy introduces a trade-off, where higher accuracy increases AoDT, leading to a lack of synchronization between the digital and physical twins. To this end, this work develops a digital twin framework that optimizes a DT-aided industrial IoT network by balancing accuracy and synchronization. We consider a system composed of several physical entities, each monitored by a group of IoT devices. UAVs are deployed to collect measurements from all IoT devices and process them to determine the status of the physical entities in the system. Our work utilizes the novel age of digital twin metric, which accounts for both upload and processing of sensitive data through employment of UAVs as edge devices to provide computing services and efficient data collection and transmission to the base station. To the best of our knowledge, no research work has tackled a similar problem with IoT devices sending fresh status updates to the digital twin platform to build a highly synchronized and accurate digital twin, with each physical entity being monitored by multiple IoT devices.   

The paper is organized as follows. We start by surveying recent relevant literature in Section~\ref{literature} and highlight our contributions. Section~\ref{system model} presents the system model and its key components. Section~\ref{Problem Formulation} formulates the problem as a mixed integer non-convex program, and Section~\ref{Solution} proposes a solution approach. Simulation results and analysis are provided in Section \ref{results}. Finally, Section~\ref{conclusion} concludes the work and presents interesting future directions.

\section{Literature Review}\label{literature}

To build a high-fidelity digital twin, various challenges are involved. According to \cite{hu2021digital,IETFDatatracker}, the main challenges include data acquisition and processing, high-fidelity modeling, and real-time communication between virtual and real twins. The precision and efficiency of data collection significantly affect the quality of the DT model \cite{hafeez2024challenges}. 
In this section, we survey selected related literature where that utilizes mobile edge computing in delay sensitive IoT applications to empower the DT technology.

Authors in \cite{Ita24} developed a high-fidelity digital twin model where its consistency with the physical system is measured through accuracy and synchronization. They deployed a UAV that follows an optimized trajectory to collect data from multiple industrial IoT devices. The data was then processed through several virtual machines equipped in the UAV to determine the status of physical entities, which was subsequently uploaded to the digital twin running at the base station. Authors introduced the age of digital twin (AoDT) metric to represent the synchronization gap between the actual status of a physical entity and its digital replica. The main aim was to develop a DT model consistent with the physical system, relying on the amount of data collected and on AoDT as crucial metrics to enhance DT accuracy and synchronization. 

In \cite{zhao2023adaptive}, authors aim to optimize the trajectories of multiple UAVs serving several IIoT devices in an industrial urban field using DT technology. The main aim of the study is to minimize the total energy consumption and improve the efficiency of UAVs in terms of latency and computation. Several IIoTs devices with different tasks; classified as delay sensitive and computational sensitive tasks, are deployed in different subareas. Each IIoT device sends its requirement state data via the UAV to the DT server, which updates the global state of the system and assigns UAVs to perform these tasks. The authors proposed dueling deep Q-networks with prioritized experience replay to optimize trajectory planning and balance latency, computation, and energy.


Li \textit{ et al.} in \cite{li2025budget} studied the problem of synchronizing the digital twin with its physical counterpart using an energy-constrained UAV that collects data from sensors deployed in a remote region. Each sensor is powered by a solar panel and its generated data is energy constrained. Authors assume that the digital twin is updated in discretized time slots. Assuming that the DTs of sensors are stored in cloudlets in a mobile edge computing network, the update budget is that the DT states of a specific group of sensors are synchronized with their physical counterparts at each time slot. The ain is to minimize the average DT state staleness of all objects given an energy-limited UAV.

Authors in \cite{li2024aoi} studied the problem of digital twin placement for improving user satisfaction based on minimum user query service delays that depends on digital twin data freshness which, in turn, relies on the age of information. They proposed two different problem scenarios: static digital twin placement and dynamic digital twin placement based on launching new digital twins periodically. Both scenarios involved deploying digital twins in cloudlets to maximize user satisfaction based on information freshness and subject to computational capacity constraints.

In \cite{kim2024real}, authors proposed a digital twin edge network where two types of real-time tasks in digital twin applications are considered: update tasks for periodic synchronization with the digital twin, and inference tasks for testing the response of a physical system by simulating the corresponding digital twin by users. The system consisted of an edge server with a set of physical systems, and multiple users. For each physical system, the corresponding cyber twin is implemented in the edge server to provide DT applications. Update tasks are generated by data collected from the physical systems while inference tasks are requested by users on demand. Authors formulated an optimization problem to maximize digital twin freshness ratio. Authors proposed a DT task scheduler that prioritizes update tasks based on the number of corresponding inference tasks. The two latter studies (\cite{li2024aoi},\cite{kim2024real}) were conducted without deploying any UAVs.

Based on the literature surveyed above and as illustrated in Table~\ref{relatedwork}, no existing research work, to the best of our knowledge, has jointly optimized accuracy and synchronization to build a multi-UAV-aided digital twin for an IoT network where multiple sensors are monitoring a single physical process. The main contributions of this work are as follows: 
\begin{enumerate}
\item We propose a digital twin framework that builds a digital twin with high accuracy and synchronization with the physical system in an IoT network. Every physical entity of the physical system is monitored by a group of IoT devices that upload their measurements to multiple UAVs optimally positioned. The UAVs process all collected measurements to determine the status of the monitored physical entities and deliver this information to the digital twin running at the base station for system monitoring and management.   
\item We derive a closed form expression for the novel AoDT metric introduced in \cite{Ita24} and used it to evaluate the synchronization difference between the physical system and its digital counterpart. This will guarantee tackling the challenge of DT real-time synchronization stated earlier. 
\item We mathematically formulate the problem as a mixed integer non-convex program to maximize the accuracy of the digital twin by optimizing the amount of collected data while ensuring a target synchronization degree with the physical system. 

\item We implement an iterative optimization approach to determine the optimized placement scheme for UAVs while maximizing data collection and ensuring freshness using successive convex approximations.
\item  We present a range of simulation results to compare the performance of our proposed approach with baseline methods and demonstrate the effectiveness of the solution to build a highly synchronized and accurate digital twin.
\end{enumerate}


\begin{table*}[ht]
\caption{Related Work Comparison}\label{relatedwork}
\centering
\begin{tabular}{|c|c|c|c|c|l|l|}
\hline
\begin{tabular}[c]{@{}l@{}}\textbf{Reference} \\\textbf{Number}\end{tabular} & \begin{tabular}[c]{@{}l@{}}\textbf{AoI} \\ \textbf{Metric} \end{tabular}  & \begin{tabular}[c]{@{}l@{}}\textbf{DT} \\ \textbf{Accuracy} \end{tabular} & \begin{tabular}[c]{@{}l@{}}\textbf{UAV(s)} \\\textbf{Deployment}\end{tabular} &  \begin{tabular}[c]{@{}l@{}}\textbf{UAV(s)}\\\textbf{Motion}\end{tabular}&\begin{tabular}[c]{@{}l@{}}\textbf{Algorithm} \\\textbf{Used}\end{tabular} & \textbf{Objective}\\
\hline

\cite{Guo23}  &  &  & Multiple & Stationary  &  \begin{tabular}[c]{@{}l@{}}Heuristic Greedy \\Deep Q-learning \end{tabular} & \begin{tabular}[c]{@{}l@{}}
DT-assisted UAV deployment \\strategy with minimum average \\delay and hybrid task offloading \end{tabular} \\
\hline

\cite{Ita24} & \checkmark & \checkmark &Single &  \checkmark &Decomposition & \begin{tabular}[c]{@{}l@{}} Propose a high fidelity DT \\model with high accuracy \\between the digital and physical \\systems achieved through maxi\\-mizing the amount of data \\collected by all IoT devices.\end{tabular}\\
\hline

\cite{li2025budget} & \checkmark &  & Single & \checkmark & \begin{tabular}[c]{@{}l@{}}Award collection maximization \\ with deep learning \end{tabular}& \begin{tabular}[c]{@{}l@{}}Optimize DT synchronization \\given energy constraints \end{tabular}\\
\hline

\cite{li2024aoi} & \checkmark & & & &\begin{tabular}[c]{@{}l@{}} \\ Decomposition with approximation   \end{tabular}& \begin{tabular}[c]{@{}l@{}}Optimize DT placement to \\enhance data freshness \end{tabular}\\ 
\hline
\cite{zhao2023adaptive} & &  & Multiple & \checkmark & \begin{tabular}[c]{@{}l@{}}Dueling deep q-learning with \\prioritized experience replay \end{tabular}& \begin{tabular}[c]{@{}l@{}}Optimize UAV trajectories in real \\ time under energy constraints \\   \end{tabular}\\
\hline
\cite{kim2024real} & \checkmark &  & &  & \begin{tabular}[c]{@{}l@{}} Status based and policy based \\algorithm \end{tabular} & \begin{tabular}[c]{@{}l@{}}Maximize DT freshness\\ \end{tabular}\\
\hline

Our Work & \checkmark & \checkmark & Multiple & Stationary & Successive convex approximation & \begin{tabular}[c]{@{}l@{}} Propose a high fidelity \\digital twin with practical settings \\optimizing UAV placement, \\accuracy and real time \\synchronization \end{tabular} \\

\hline
\end{tabular}
\end{table*}

\section{System Model}\label{system model}

\begin{figure}[ht]
  \centering
  \includegraphics[width=\columnwidth]{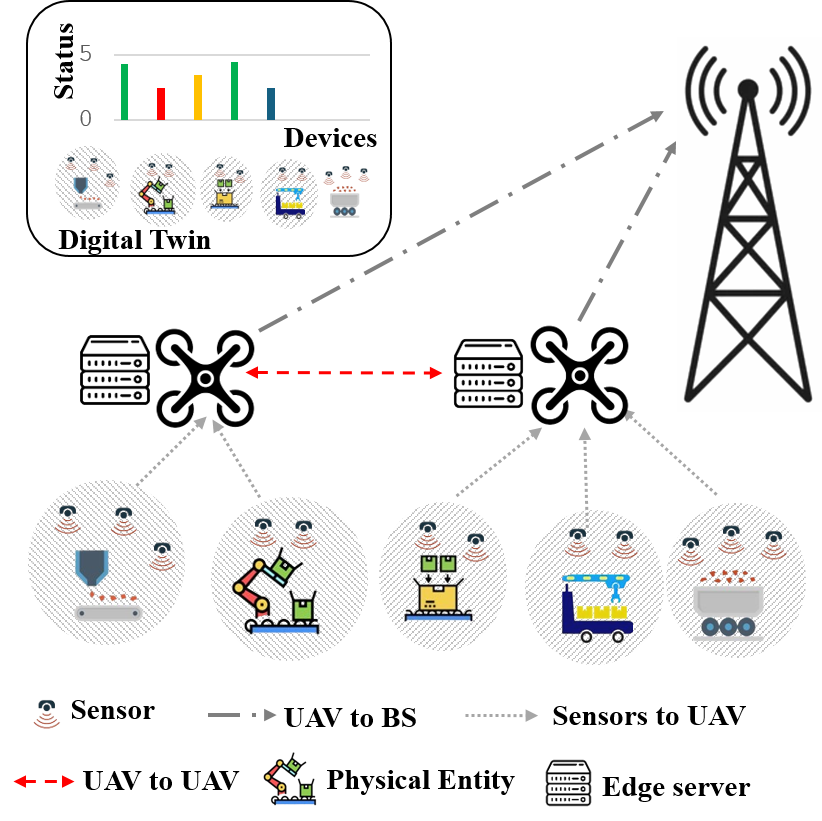}
  \caption{System Model}
  \label{fig:System_Model}
\end{figure}

We consider an industrial facility equipped with a digital twin running at a nearby base station located at $(x_{BS}, y_{BS}, z_{BS})$. A set of $I$ IoT devices, denoted by $\mathcal{I}=\{1,2,...,I\}$, is deployed in the facility at ground level at $\{x_{i}, y_{i}, 0\}$ for monitoring various aspects of the physical environment. IoT devices are supposed to collect real-time measurements, such as temperature, pressure, humidity, vibration, flow rate, and others, of the components of the industrial system, named physical entities. Due to their limited transmission power and lack of processing capabilities, IoT devices deliver their readings to a set of $J$ stationary UAVs denoted by $\mathcal{J}=\{1,2,...,J\}$, and that are optimally positioned at the coordinates $(x_{j}, y_{j}, H)$, $\forall j\in \mathcal{J}$ where $H$ is the height UAVs are positioned. The UAVs are assumed to be fully meshed with a high-speed link and they use their computational power to process the measurements and determine the current status of the physical entities. Status updates of the physical entities are then uploaded to the digital twin on the BS in a way to ensure high synchronization with the physical system. Fig.~\ref{fig:System_Model} illustrates this system model.  

Depending on the size, structure, or functionality of each physical entity, multiple IoT devices may be needed to capture measurements from different angles or locations. Each IoT device $i$ generates tasks of size $S_i$ in bytes that are uploaded to UAV $j$. However, if UAV $j$ lacks sufficient computational resources to process a task, it forwards the task to another UAV $k$ for processing. UAVs are considered to operate in full duplex mode, allowing them to receive and transmit simultaneously, while using orthogonal channels to avoid self-interference. Moreover, UAVs opt to use the most recent measurements coming from IoTs to ensure that the resulting status reliably mirrors the current status of the monitored physical entity. This being said, a UAV discards old data once it obtains fresh measurements from a respective IoT.

\subsection{Communication Channel Model}\label{Communication}


Ground-to-air (G2A) communication between UAVs and ground IoT devices is based on a probabilistic path loss model that accounts for line-of-sight (LoS) and non-line-of-sight (NLoS) components. Accordingly, we employ the following LoS and NLoS channel models between IIoT device $i$ and UAV $j$~\cite{Guo23}:
\begin{equation}
L^{LoS}_{i,j} = J_{FS} + 20 \log d_{i,j} + \eta_{LoS},
\end{equation}
and
\begin{equation}
L^{NLoS}_{i,j} = J_{FS} + 20 \log d_{i,j} + \eta_{NLoS},
\end{equation}
where J$_{FS} = 20 \log f_c + 20 \log \frac{4 \pi}{c}$, the variable \( f_c \) signifies the system's carrier frequency, and \( c \) is the speed of light's constant. The parameters \( \eta_{LoS} \) and \( \eta_{NLoS} \) are additional attenuation factors specific to the line-of-sight (LoS) and non-line-of-sight (NLoS) link environments, respectively. $d_{i,j}$ denotes the Euclidean distance between device $i$ and UAV $j$ is represented by: 
\begin{equation}\label{equation1}
    d_{i,j}= \sqrt{(x_i - x_j)^2 + (y_i - y_j)^2+ H^2}.
\end{equation}

In addition, the probability of a LoS communication link, denoted as \( P^{LoS}_{i,j} \), between a UAV $j$ an IoT $i$ device is represented as: 
\begin{equation}\label{equation5}
P^{LoS}_{i,j} = \frac{1}{1 + a \cdot \exp(-b \cdot (\arcsin(\frac{z_{j}}{d_{i,j}}) - a))}
\end{equation}
where $a$ and $b$ are environment parameters. As a result, the average path loss is expressed to be:
\begin{equation}\label{equation6}
L^{avg}_{i,j} = P^{LoS}_{i,j} L^{LoS}_{ij} + (1 - P^{LoS}_{i,j}) L^{NLoS}_{i,j}
\end{equation}

Based on the above, the achieved uplink rate from IIoT $i$ to UAV $j$ is computed as \cite{Guo23}:
\begin{equation}\label{rate}
r_{i,j} = B_{ij} \log_2\left(1 + p_i \cdot 10^{\frac{-L^{avg}_{i,j}}{10}} \cdot \frac{1}{\sigma^2}\right)
\end{equation}
where $B_{ij}$ represents the channel bandwidth that is distributed among the devices based on their needs, $p_i$ is the power level of IoT device $i$, and ${\sigma^2}$ is the Gaussian white noise power.

For simplicity, we assume a high-speed LoS link between UAVs. Therefore, the elapsed time to upload data from the associated UAV $j$ with the IoT device $i$ to another UAV $k$ is considered constant and equal to $T_{u2u}$. 
We also consider that the base station is connected to the core network via a wired fiber link, while each UAV communicates with the base station via a strong line of sight wireless backhaul link with high communication speed and low latency \cite{ji2023trajectory}. 
According to related literature~\cite{Han21}, the download time of the computed results from the UAV to the BS can be neglected, since the processed data is relatively small.

\subsection{Data Generation and Computing Models} \label{processing}

During monitoring a physical entity, each IoT device $i$ generates data based on a Poisson distribution with an average rate $\lambda_i$. The data generated from multiple IoT devices is transferred to UAV $j$ for potential processing.

Since each associated IoT device generates its tasks according to a Poisson process with rate $\lambda_i$, and assuming a constant propagation delay from the IoT device $i$ to UAV $j$, the aggregated workload from a set $\mathcal{N}_j$ of IoT devices on UAV $j$ can also be modeled as a Poisson process with rate $\sum_{k \in \mathcal{N}_j}\lambda_k$. 

We assume that the service times at UAV $j$ are independent and identically distributed exponential random variables, with an average service time of $\frac{1}{\mu_j}$, where ${\mu_j}$ resembles the average service rate of the UAV $j$. The service rate ${\mu_j}$, measured in requests per second, can be expressed as:
\begin{equation}
    \mu_j = \frac{f_j}{L}
\end{equation}
where $f_j$ is the processing capacity of UAV $j$ in cycles per second, and L is the average task size in cycles.
Hence, we model each UAV $j$ as an M / M / 1 queueing system with an arrival rate $\lambda_{total,j} = \sum_{k \in \mathcal{N}_j}\lambda_k$ and a service rate $\mu_j$~\cite{Ala19,Tou23}.

We define the offered load $\rho_i$ from IoT device $i$ to the processing UAV $j$ by:
\begin{equation}
    \rho_i = \frac{\lambda_i}{\mu_j}
\end{equation}
Thus, the total offered load at UAV $j$ where the tasks from a set of $\mathcal{N}_j$ IoT devices are processed is:
\begin{equation}
    \rho_{total,j} = \frac{\lambda_{total,j}}{\mu_j}= \frac{\displaystyle\sum_{k \in \mathcal{N}_j}\lambda_k}{\mu_j}
\end{equation}

\subsection{Modeling Quality of Service in a Digital Twin} \label{AoI}

Many research for defining methods 
and metrics to evaluate digital twins is emerging. Authors in \cite{khan2023correspondence} illustrated that ensuring real-time synchronization between a digital twin and its physical counterpart and maintaining high accuracy in a digital twin is crucial in effective digital twin adoption for diiferent applications. Authors in \cite{bellavista2024odte} introduced digital twin entanglement metric that represents to which extent
 the digital and physical systems' synchronization process fulfills the needs of a specific application. Authors took into account the freshness of the collected data and the ratio of collected to total data. This ratio reflects the DT accuracy. Moreover, authors in \cite{kim2023real} illustrated that DT freshness may be defined using the age of information concept as the elapsed time after a DT has been updated. DT Freshness can be best represented through a novel metric introduced in \cite{Ita24} as age of digital twin metric (AoDT). Authors in \cite{Ita24} considered a simple scenario where one Iot monitors one physical process and proposed a problem to optimize DT accuracy and freshness. However; in practice, multiple Iot devices might monitor the same physical entity. In this work, we formulate this closed form expression to model age of digital twin in a more practical scenario.

The age of information (AoI) is a destination-centric measure for information freshness that counts the time that has passed since the last fresh update was generated at the source. In this study, we apply the concept of AoI to the setting of digital twins, where we need to express the freshness of the physical entity's status through the responsive DT. As a result, we utilize the Age of Digital Twin (AoDT) metric introduced in \cite{Ita24} and defined as the elapsed time between data generation and the status update received by the BS.

The calculation of the AoDT for each monitored process 
 accounts for different phases after data collection, namely:  
\begin{itemize}
    \item Upload phase:  The data is transmitted to the associated UAV through the wireless channel for processing, and then forwarded from the associated UAV to another one when the first UAV lacks computation resources.
    \item Processing phase: The uploaded data enters the computation queue at the UAV and is then processed to obtain results.
    \item Download phase: The data is processed from the UAV to the BS via the LoS channel to update the DT's status.
\end{itemize}

\begin{figure}[tb!]
	\centering
	\includegraphics[width=0.45\textwidth]{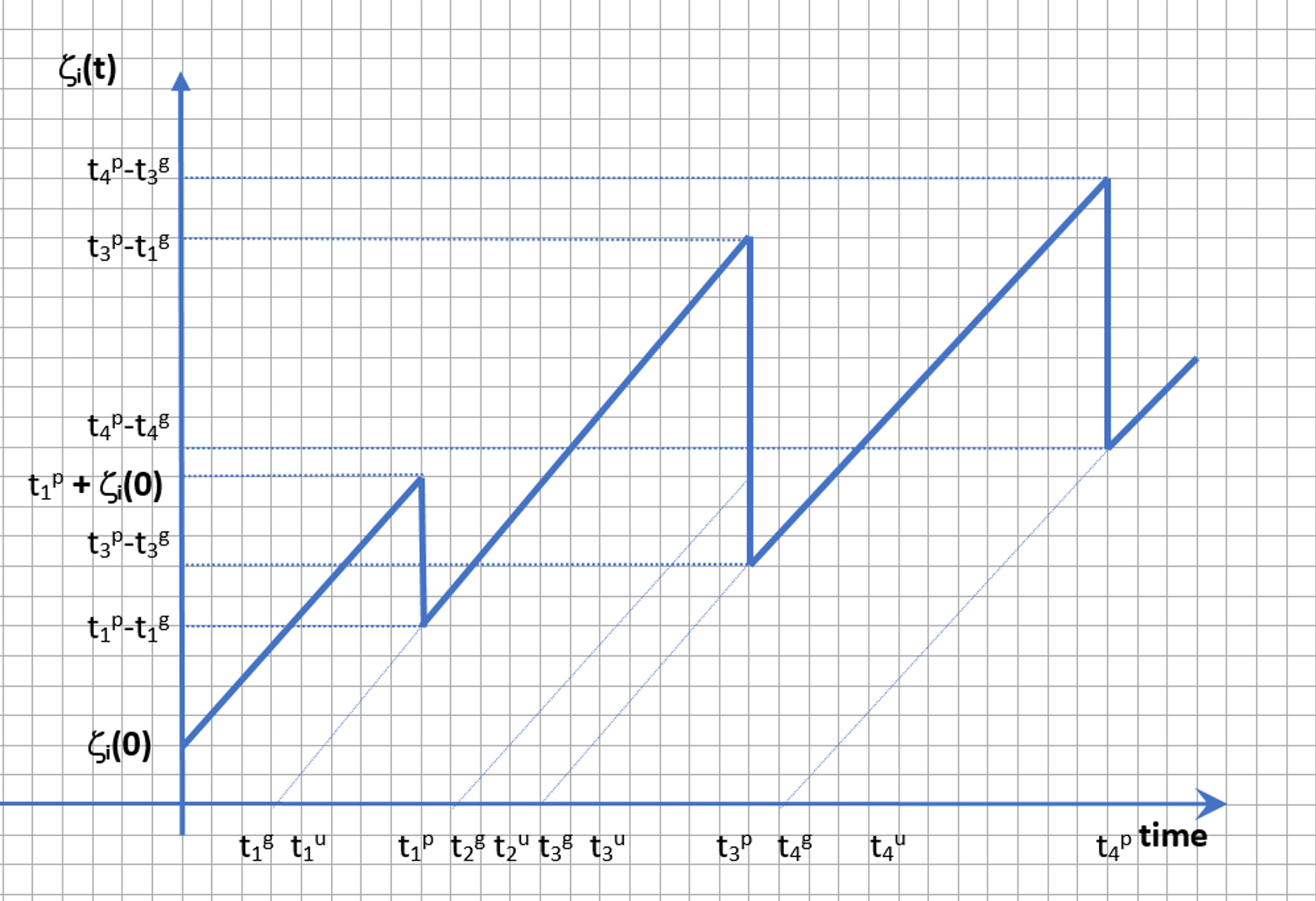}
	\caption{Evolution of AoDT example}
	\label{fig:AoDT}
\end{figure}
We define the instantaneous AoDT at time $t$ for the ith IoT device as follows:
\begin{equation}\label{eq:AoDT}
    \zeta_i(t) = t - u_i(t)
\end{equation}
where $u_i(t)$ is the instant generation of the last update of the DT from IoT $i$.
Figure.~\ref{fig:AoDT} illustrates an example of the evolution of the AoDT $\zeta(t)$ for IoT device $i$ over time $t$. Without loss of generality, suppose we start monitoring at time $t=0$, when the computation queue is empty and the age is $\zeta(0)$.  In the absence of updates, the AoDT for the DT corresponding to the source $i$ climbs linearly with time and is reset to a lower value when an update is received. 

Update $l$ for IoT device $i$, generated at time $t^g_l$, is uploaded to the associated UAV $j$ and then to another UAV $k$ if the UAV $j$ lacks computation resources. We denote by $t^u_l$ the upload time. 
As a result, we can express the upload duration $X_l = t^u_l - t^g_l$ of task $l$ for IoT device $i$ as follows:
\begin{equation}
    X_l = 
    \begin{cases}
    \frac{S_i}{r_{ij}} & \text{if the task } l \text{ is processed }\\
    & \text{ by the associated UAV} j\\
    \frac{S_i}{r_{ij}} + T_{u2u}  & \text{otherwise}
    \end{cases}
    \label{eq:upload}
\end{equation}
where $S_i$ represents the generated task size in bytes from IoT device $i$, and $T_{u2u}$ is the required time to upload the generated data of the device $i$ from its associated UAV $j$ to another one, when $j$ lacks sufficient computation resources. 

Assuming stable radio conditions for each IoT device $i$ with tasks of the same size $S_i$, the upload time $X_l$ is constant and equal to $D_i$ for all $l$. 

After the uploading phase, data is processed at $t^p_l$, and the result is received by the BS to update the status of the corresponding DT at $t^d_l$. At $t^d_l$, the AoDT $\zeta_i(t^d_l)$ is reset to ($t^d_l-t^g_l$).

Regarding the download phase of the processed task $l$ from UAVs to the BS, the download duration is given by: 
\begin{equation}
    Z_l = t^d_l - t^p_l
\end{equation}

Therefore, the evolution of the AoDT is mainly affected by the processing phase since the time elapsed at other phases is considered constant. Given that we have modeled every UAV $j$ as a M/M/1 system and infinite buffer, we can select either first-come, first-serve (FCFS) or last-come, first-serve (LCFS) queuing discipline. 
Noting that the FCFS queue paradigm allows new update messages to be queued behind outdated messages issued earlier, and that we want to transfer the most recent data from IoT to reduce the AoDT, it is preferable to use the LCFS queuing discipline. Considering LCFS, fresh information from an IoT device preempts any previously queued update packets, and the preempted data is deleted. Furthermore, based on the performance evaluation in~\cite{Yat19}, the AoI values achieved using the LCFS strategy outperformed those obtained under the FCFS strategy. Hence, we will consider LCFS with preemption in service (LCFS-S). When a new packet is uploaded, it preempts the packets that are being served. We can show from Fig.~\ref{fig:AoDT} after the third generated update task at time $t^g_3$ is uploaded to the UAV at time $t^u_3$, it preempts the previous task in service to process recent data.

For the proposed system, multiple IoT devices monitor a single physical process. Each device $i \in N_k$, where $N_k$ represents the set of devices monitoring the same process, generates updates at a rate $\lambda_i$ following a Poisson process. Updates are transmitted directly to a UAV without an aggregator. Then, the UAV processes each update and updates the state of the corresponding Digital Twin (DT). Once an update is processed, the Age of Digital Twin (AoDT) for the corresponding DT is reset.

Since multiple IoT devices directly send updates to the UAV, the min rate should be taken into consideration to wait for the slowest update to be received, $\lambda_{N_k}$ is given as:
\begin{equation}
    \lambda_{N_k} = \min_{i \in \mathcal{N}_k}  \lambda_i,
\end{equation}

The Average Age of Information (AAoI) for a multiple-source, single-server system with LCFS preemption in service \cite{Yat19} is:
\begin{equation}
    \Delta = \frac{1}{\lambda} \left( 1 + \frac{\lambda}{\mu} \right),
\end{equation}
where $\lambda$ is the total arrival rate and $\mu$ is the service rate.

For our system, the average arrival rate $\lambda_{N_k}$ replaces $\lambda$. Hence, the average AoDT for a Digital Twin $DT_k$ monitoring the physical process is:
\begin{equation}
    \Delta_k = D_{N_k} + \frac{1}{\lambda_{N_k}} \left( 1 + \frac{\lambda_{N_k}}{\mu} \right),
\end{equation}
where $D_{N_k}$ represents the maximum upload time for updates from the devices in $N_k$ and is given as:

\begin{equation}
    D_{N_k} = \max_{i \in N_k} D_i,
\end{equation}
where $D_i$ is the upload time for device $i$.

Substituting $D_{N_k}$ into the AoDT expression, we obtain the final form of the average AoDT:
\begin{equation}
    \Delta_{DT_k} = \max_{i \in N_k} D_i + \frac{1}{\lambda_{N_k}} \left( 1 + \frac{ \sum_{i \in N_k} \lambda_i}{\mu} \right).
\end{equation}

\section{Problem Formulation}\label{Problem Formulation}
In this section, we formulate a non-convex optimization problem ($\mathcal{P}$) to maximize the DT accuracy. We consider the scenario in which multiple IoT devices monitor a single physical entity/process. 
Each IIoT device $i \in \mathcal{I}$ is associated with a single UAV $j \in \mathcal{J}$ that receives uploaded tasks and processes them. However, when the associated UAV has limited computational capabilities, it will delegate the tasks to another UAV for processing. 
We define three binary decision variables $a_{ij}$, $b_{ij}$ and $\delta_{ik}$. The first variable $a_{ij}$ indicates whether IIoT $i$ is associated with UAV $j$.
\begin{equation}
a_{ij} =
\begin{cases} 
1 & \text{if IoT device } i \text{ is associated with UAV} j \\
0 & \text{otherwise}
\end{cases}
\end{equation}
The second binary variable $b_{ij}$ indicates whether the tasks of IoT $i$ are processed by UAV $j$.
\begin{equation}
b_{ij} =
\begin{cases} 
1 & \text{if the tasks generated by IoT device } i \\ & \text{ are processed by UAV} j\\
0 & \text{otherwise}
\end{cases}
\end{equation}
Since multiple sensors are monitoring one process, we define $\mathcal{K}$ groups or sets of IoT sensors where each set ${N_k}$  is monitoring a process as stated earlier. 

The third binary variable $\delta_{ik}$ indicates the set to which the IoT device belongs. 
\[
 \delta_{ik} =
\begin{cases}
1, & \text{if IoT device } i \text{ is assigned to a set } k \\
0, & \text{otherwise}
\end{cases}
\]
The formulated problem ($\mathcal{P}$) determines the optimized position for each UAV $j \in \mathcal{J}$. Furthermore,  ($\mathcal{P}$) derives for each IIoT device $i \in \mathcal{I}$ its associated UAV and the UAV responsible for processing its tasks.

\begin{align}
    & (\mathcal{P}): & & \max \sum_{i=1}^{I} \sum_{j=1}^{J} a_{ij} r_{ij} \label{obj} \\
    & \text{s.t.} 
    & & \sum_{j=1}^{J} a_{ij} = 1, \quad \forall i \in  N_k, \label{ctr1} \\
    & & & \sum_{j=1}^{J} b_{ij}^k = \delta_{ik}, \quad \forall i \in N_k, k \in \mathcal{K}, \label{ctr2} \\
    & & & \sum_{k=1}^{K} \delta_{ik} = 1, \quad \forall i \in {N_k}. \label{ctr11} \\
    & & & \delta_{ik} . \delta_{nk} . b_{ij}^k = \delta_{ik} . \delta_{nk} . b_{nj}^k, \quad \forall i, n \in N_k, j \in \mathcal{J},  k \in \mathcal{K},  \label{ctr9} \\
    & & & \frac{f_j}{L} - \sum_{i=k}^{K}\sum_{i=1}^{I} b_{ij}^k \lambda_i \geq 0, \quad \forall j \in \mathcal{J}, \label{ctr3} \\
    & & & r_{ij} \geq a_{ij} R_{\text{min}}, \quad \forall i \in {N_k}, j \in \mathcal{J}, \label{ctr4} \\
    & & & B_{ij} \leq a_{ij} \cdot M, \quad \forall i \in {N_k}, j \in \mathcal{J},M \to \infty  \label{ctr5} \\
    & & & \sum_{i=1}^{I} \sum_{j=1}^{J} B_{ij} \leq B_{\text{sys}}, \label{ctr6} \\
    & & & \sqrt{(x_j - x_l)^2 + (y_j - y_l)^2 + (z_j - z_l)^2} \geq \theta, \quad \forall j, l \in \mathcal{J}, \label{ctr7} \\
    & & & \lambda_{N_k} \leq \lambda_i \quad \forall i \in N_k  \label{ctr13} \\
    & & & \mathrm{v}^{k} \leq T_k \quad \forall k \in \mathcal{K}, \label{ctr12} \\
    & & & \mathrm{v}^{k} \geq \sum_{j=1}^{J}a_{ij}\frac{S_i}{r_{ij}}  + \frac{1}{\lambda_{N_k}} 
    \left( 1 + \frac{ \sum_{i \in N_k} \lambda_i}{\mu} \right) \nonumber \\
    & & & \quad +  \left( 1 - \sum_{j=1}^J a_{ij} b_{ij}^k \right) \cdot T_{\text{u2u}}
    \quad \forall i \in N_k, k \in \mathcal{K}, \label{ctr8} 
\end{align}

Problem $(\mathcal{P})$ optimizes data collection from IoT devices by maximizing the sum rate while considering constraints related to computation tasks, AoDT, available resources, and overall system performance. The objective function~(\ref{obj}) aims to maximize the sum rate. Constraints~(\ref{ctr1}) and~(\ref{ctr2}) ensure that each IoT device $i$ is associated with a single UAV $j$ and that the tasks of each device $i$ are being processed by a single UAV $j$. Constraint ~(\ref{ctr11}) ensures that each IoT device belongs to one set only. Constraint~(\ref{ctr3}) is crucial to maintaining a stable queuing system where we ensure that the rate at which computational requests arrive is lower than the service rate for each UAV.  To guarantee a minimum quality of service (QoS), constraint~(\ref{ctr4}) ensures that the communication rate between the IoT devices and the UAV is higher than the target value.  Constraint~(\ref{ctr5}) ensures that the bandwidth is set to zero if there is no association between IoT $i$ and UAV $j$. Constraint~(\ref{ctr6}) ensures that the total uplink bandwidth of the system does not exceed the available system bandwidth of value $ B_{sys}$. Constraint~(\ref{ctr7}) mandates a minimum safety distance $\theta$ between any two deployed UAVs. 
Constraints~(\ref{ctr12}) and~(\ref{ctr8}) ensure that the average AoDT $\Delta_k$ for IoT device $i$ remains below a predefined threshold $T_K$ to maintain the freshness of data updates. 

\section{Solution Approach}\label{Solution}
In our simulation, we aim to solve the problem ($\mathcal{P}$) which is a mixed integer non-convex optimization problem. 
Our solution approach addresses the difficulty of this problem by linearizing and convexifying the constraints. In our solution, we propose a successive convex approximation approach to determine the optimized positions of the stationary UAVs to maximize the data collection rates under AoDT threshold to improve both accuracy and synchronization of the DT.
We encounter several nonlinear constraints in our problem like (\ref{equation1}), (\ref{equation5}), (\ref{equation6}) and others. Some of these constraints are linearized using Taylor series expansion, and others using the auxiliary variables approach.
Algorithm(\ref{algorithm1}) presents the successive convex approximations of ($\mathcal{P}$).
\begin{algorithm}
\caption{Solution of Problem $\mathcal{P}$}
\label{algorithm1}
\begin{algorithmic}[1]
\State \textbf{Input:} $\mathcal{I}$, $\mathcal{J}$, positions of IoT devices, position of the base station, error tolerance
\State \textbf{Initialization:} Set initial position for UAVs, bandwidth, and iteration number $n$
\While{$|\text{Obj}\mathcal{P}(n) - \text{Obj}\mathcal{P}(n-1)| \leq \epsilon$}
    \State Solve the convex problem $\mathcal{P}$ to obtain optimal UAV positions and system bandwidth
    \State Update UAV positions
    \State Update bandwidth value
    \State Update sum rate matrix
    \State Update $n = n + 1$
\EndWhile
\State \textbf{Output:} The optimal UAV positioning and the maximum system bandwidth
\end{algorithmic}
\end{algorithm}

We used the Matlab Mosek cvx solver, which applies a primal-dual interior-point algorithm that simultaneously solves the primal and dual problems to achieve good accuracy. This solver iteratively solves continuous convex optimization problems by moving within the feasible region while maximizing the objective function. To keep variables within the feasible region, interior-point methods work by using Newton’s method on "barrier" functions. Also, Mosek's primal-dual interior-point method ensures fast convergence for the high complexity of the problem involving UAV placement and data collection. Thus, the worst-case complexity of this problem can be roughly estimated as C $\sim \sqrt{\text{number of variables}}$ because the complexity depends on the number of Newton steps, in which the latter is influenced by the number of variables of the optimization problem and the iterations needed for the algorithm to converge given an initial start as stated in ~\cite{Ita24}.

\section{Simulation Results and Performance Analysis}\label{results}

In this section, we evaluate the proposed solution approach formulated above. We consider an industrial field of $500 \times 500 \, \text{m}^2$ where ten IoT devices are randomly deployed, with every five devices monitoring one process. We consider the system parameters as shown in Table~\ref{table:simulation_parameters}. These parameters are assumed unless otherwise stated.  We present different simulation results of the optimized solution compared to two baseline approaches, k-means and random placement of UAVs as an initialization step. The k-means positioning approach determines UAV locations by minimizing the sum of distances between the IoT devices and their cluster centroids where the UAV is supposed to be positioned. In contrast, the random deployment method assigns UAV positions without any specific strategy, disregarding the locations of IoT devices or any optimization concerns. 
\begin{table}[h!]
\centering
\begin{tabular}{|c|c|}
\hline
\textbf{Parameter} & \textbf{Value} \\ \hline
Minimum data rate, $R_{\text{min}}$ & 10000 bits/sec \\ \hline
Minimum bandwidth allocation, $B_{\text{min}}$ & 20000 Hz \\ \hline
Distance limit between UAVs, $\theta$ & 10 m \\ \hline
UAV altitude, $z_{\text{UAV}}$ & 100 m \\ \hline
Upload time between UAVs, $T_{\text{u2u}}$ & 0.3 sec \\ \hline
Carrier frequency, $f_c$ & $1 \times 10^6$ Hz \\ \hline
Task arrival rate, $\lambda_i$ & $2$ tasks/sec \\ \hline
UAV processing frequency, $f_j$ & $2 \times 10^8$ Hz \\ \hline
Speed of light, $c$ & $3 \times 10^8$ m/s \\ \hline
AoDT threshold, $T_K$ & $2.8$ sec \\ \hline
LoS additional loss, $\eta_{\text{LoS}}$ & 1 (linear scale) \\ \hline
NLoS additional loss, $\eta_{\text{NLoS}}$ & 21 (linear scale) \\ \hline
Noise power, $\sigma$ & $10 \times 10^{-3}$ W \\ \hline
Power, ${\text{pi}}$ & 0.2 W \\ \hline
Environment parameter, $a$ & 9.61 \\ \hline
Environment parameter, $b$ & 0.16 \\ \hline
\end{tabular}
\vspace{5pt}
\caption{Simulation parameters}
\label{table:simulation_parameters}
\end{table}
For the simulation, we used the DELL laptop 13th Gen Intel(R) Core(TM) i7-1355U, an x64-based processor, and a CPU core running at 2.06 GHz with 16 GB RAM. Note that each data point in the figures represents the average of 20 different random runs.
\begin{figure}[ht]
  \centering
  \includegraphics[width=1\columnwidth]{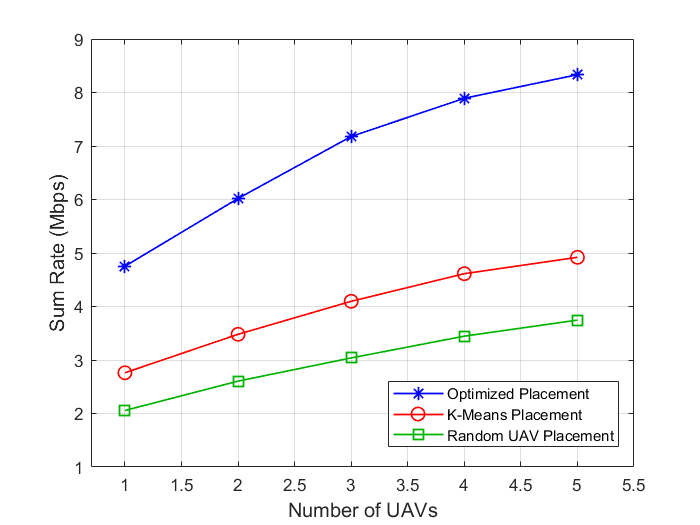}
  \caption{Sum rate in terms of UAVs number where $\mathcal{I}=10$}
  \label{fig:Figure4}
\end{figure}
We varied multiple metrics to plot and compare the sum rate values to analyze the performance of the proposed solution. 
Fig.~\ref{fig:Figure4} shows the relationship between the number of deployed UAVs and the sum rate. The sum rate improves across all three approaches as more UAVs are deployed due to enhanced resource utilization and spatial coverage, enabling efficient data collection to ensure fresh status updates for the DT. Optimized placement consistently outperforms both baseline approaches, achieving the highest sum rate of approximately 8.8 Mbps with the deployment of five UAVs, compared to 5.6 Mbps for k-means placement and 3.4 Mbps for random placement.
These results illustrate that optimized placement ensures timely updates thus improved synchronization for the digital twin, which, in turn, enhances data freshness and accuracy of the digital model.
\begin{figure}[ht]
  \centering
  \includegraphics[width=1\columnwidth]{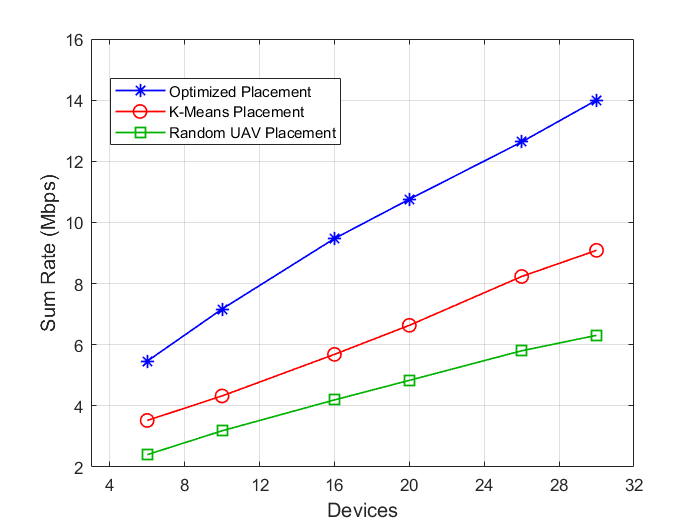}
  \caption{Sum rate in terms of IoT device count where $\mathcal{J}=3$}
  \label{fig:Figure5}
\end{figure}
Increasing the number of IoT devices, Fig~\ref{fig:Figure5} shows an increase in the sum rate for a fixed number of deployed UAVs (three), across all placement strategies. This reflects the increased data generation effect and improved utilization of network resources for parallel and efficient data collection by UAVs. Our proposed solution achieved a maximum sum rate exceeding approximately 12 Mbps for 32 devices, while the k-means and random UAV placement achieved approximately 7.8 Mbps and 5.2 Mbps, respectively. The optimized placement consistently outperforms the other methods, achieving the highest sum rate due to better spatial distribution of UAVs.
This considerable performance gap highlights the efficiency of our optimized UAV placement in ensuring optimal network coverage, balanced resource allocation, and minimal interference.
. 

\begin{figure}[ht]
  \centering
  \includegraphics[width=1\columnwidth]{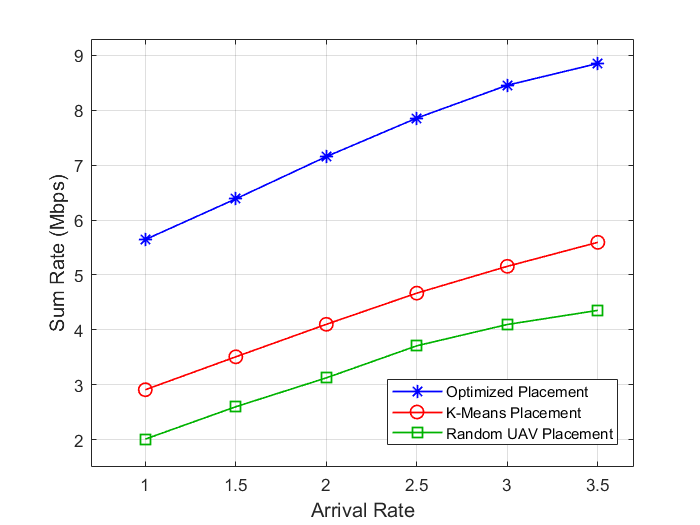}
  \caption{Sum rate in terms of tasks arrival rate, with $\mathcal{I}=10$ and $\mathcal{J}=3$}
  \label{fig:Figure6}
\end{figure}

In Fig~\ref{fig:Figure6} we increase the task arrival rate from 1 to 3.5 tasks per second and noticed that the achieved sum rate improves significantly across all UAV placement methods. However, our optimized placement strategy consistently maintains the highest sum rate of 8.9 Mbps, compared to approximately 5.3 Mbps for k-means placement and 4 Mbps for the random placement scenario. These results demonstrate how a higher arrival rate affects the total sum rate, as more data packets are transmitted.
\begin{figure}[ht]
  \centering
  \includegraphics[width=1\columnwidth]{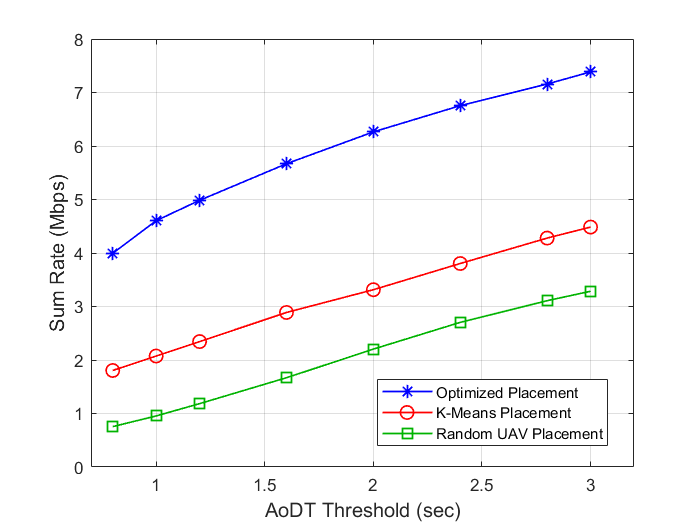}
  \caption{Sum rate in terms of age of digital twin threshold, with $\mathcal{I}=10$ and $\mathcal{J}=3$}
  \label{fig:Figure7}
\end{figure}
To better demonstrate the impact of the AoDT metric on the system performance, we compared the total sum rate in Fig~\ref{fig:Figure7} while varying the AoDT threshold, starting from 0.8 seconds, across different UAV placement strategies. All placement approaches demonstrate an increase in sum rate with a more relaxed AoDT threshold. Remarkably, the optimized placement achieved the highest performance, reaching 7.8 Mbps when the threshold was relaxed to 3 seconds, compared to k-means and random placement, which reached around 5.4 Mbps and 3.7 Mbps, respectively. 
The performance of the baseline approaches highlights their limitations in improving system efficiency due to poor spatial configuration, resulting in the under utilization of available resources. In contrast, the optimized placement significantly enhances system efficiency and network throughput, particularly as the AoDT threshold is relaxed, allowing UAVs to allocate communication resources more effectively and flexibly. 

\begin{figure}[ht]
  \centering
  \includegraphics[width=1\columnwidth]{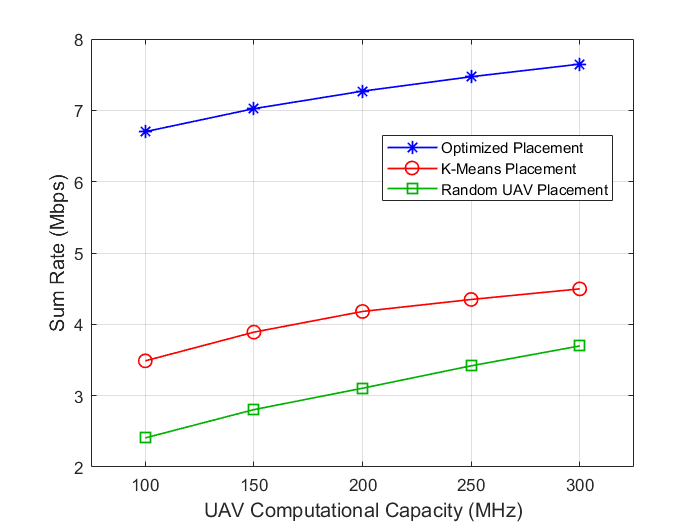}
  \caption{Sum rate in terms of UAV's computational capacity, with $\mathcal{I}=10$ and $\mathcal{J}=3$}
  \label{fig:Figure8}
\end{figure}

The results in Fig.~\ref{fig:Figure8} illustrate how sum rate performance is evaluated across different UAV computational capacities for the three placement strategies stated earlier. The figure shows that increasing UAV computational capacity improves the sum rate performance among all three plots, demonstrating that enhanced processing capabilities enable faster handling of computational tasks, which, in turn, optimizes communication resources, allowing for additional data transmission tasks. The optimized placement approach is still recording the highest sum rate, reaching approximately 7.5 Mbps at 250 MHz. In comparison, the k-means method attained a sum rate of 4.7 Mbps, while random placement reached approximately 3.3 Mbps.

\section{Conclusion}\label{conclusion}
DT technology serves as a vital tool in real-time network monitoring and performance optimization \cite{li2022digital}. To ensure efficient performance, the digital twin should predict the network state accurately and frequently. In this work, we propose a problem where a digital twin is used for monitoring an industrial field using a UAV-assisted group of IIoT devices. We consider a practical scenario where a group of multiple IoT devices monitors the same physical process. Different groups of IoT devices collect status data and upload it to multiple UAVs optimally positioned. The UAVs process all collected measurements to determine the status of the monitored physical system and deliver this information to the digital twin running at the base station for system monitoring and management. We formulate a mixed-integer non-convex program to maximize the total amount of data collected from all IoT devices while ensuring a constrained AoDT that reflects digital twin data freshness. We conduct extensive simulations and compare the results with baseline approaches including k-means and random UAV placement to demonstrate the effectiveness of the proposed solution. Some interesting future considerations include adopting the AoDT metric in different contexts where specific requirements are needed.  Moreover, we might consider deriving an age of digital twin model for hierarchical edge computing systems with associated hierarchical digital twin systems.

\end{document}